\documentclass[conference]{ieeeconf}
\IEEEoverridecommandlockouts
\usepackage{cite}
\usepackage{amsmath,amssymb,amsfonts,amsthm}
\usepackage{graphicx}
\usepackage{textcomp}
\usepackage{xcolor}
\usepackage{algorithm}
\usepackage[noend]{algpseudocode}
\usepackage{comment}
\usepackage{bm}
\usepackage[T1]{fontenc}
\let\labelindent\relax
\usepackage{enumitem}

\newtheorem{definition}{Definition}
\newtheorem{theorem}{Theorem}

\newtheorem{proposition}[theorem]{Proposition}

\DeclareMathOperator*{\argmax}{arg\,max}
\DeclareMathOperator*{\argmin}{arg\,min}

\def\BibTeX{{\rm B\kern-.05em{\sc i\kern-.025em b}\kern-.08em
    T\kern-.1667em\lower.7ex\hbox{E}\kern-.125emX}}

\title{\LARGE \bf Playing with Peaks: A Game-Theoretic Comparison of Electricity Pricing Mechanisms}

\author{Vade Shah and Jason R. Marden
\thanks{This paper is partially supported by the California Energy Commission (project \#EPIC-24-012), the NSF GRFP (grant \#2139319), and the AFOSR (grant \#FA9550-25-1-0245).}
\thanks{V. Shah ({\tt\small vade@ucsb.edu}) and J. R. Marden are with the Department of Electrical and Computer Engineering at the University of California, Santa Barbara, CA.}%
}

\begin{document}

\maketitle 

\begin{abstract}
As electricity consumption grows, reducing peak demand—the maximum load on the grid—has become critical for preventing infrastructure strain and blackouts. Pricing mechanisms that incentivize consumers with flexible loads to shift consumption away from high-demand periods have emerged as effective tools, yet different mechanisms are used in practice with unclear relative performance. This work compares two widely implemented approaches: \emph{anytime peak pricing} (AP), where consumers pay for their individual maximum consumption, and \emph{coincident peak pricing} (CP), where consumers pay for their consumption during the system-wide peak period. To compare these mechanisms, we model the electricity market as a strategic game and characterize the peak demand in equilibrium under both AP and CP. Our main result demonstrates that with perfect information, equilibrium peak demand under CP never exceeds that under AP; on the other hand, with imperfect information, the coordination introduced by CP can backfire and induce larger equilibrium peaks than AP. These findings demonstrate that potential gains from coupling users' costs (as done in CP) must be weighed against these miscoordination risks. We conclude with preliminary results indicating that progressive demand cost structures—rather than per-unit charges—may mitigate these risks while preserving coordination benefits, achieving desirable performance in both deterministic and stochastic settings.

\end{abstract}

\section{Introduction}

Global electricity consumption has grown significantly over the past decade and is expected to continue rising with the expansion of data centers, electric vehicles, and other emerging technologies \cite{IEA2025GlobalEnergyReview}. While this growth provides many environmental, financial, and societal benefits, it also poses significant new challenges for power grid operation \cite{IEA2025GridCongestion, finn2014demand}. A major challenge is that of managing \emph{peak demand}, which is defined as the maximum load the electricity grid experiences at any given time. Electricity consumption patterns are highly time-varying, and during high activity periods, massive demand can cause inefficient operation, accelerate equipment degradation, and, in severe cases, trigger blackouts \cite{EIA2020HourlyElectricityConsumption}. As electricity needs have grown, so has the intensity of these peak periods, making peak demand reduction an increasingly critical problem for the grid \cite{WilsonZimmerman2023NationalLoadGrowth}.

In recent years, \emph{pricing} has emerged as an effective tool for reducing peak electricity demand by incentivizing users with flexible electricity needs to shift usage to off-peak periods \cite{qian2013demand, muratori2015residential, dutta2017literature}. As a growing share of electricity consumption is becoming flexible due to renewable and battery technologies \cite{EIA2025ShortTermEnergyOutlook}, pricing may play an even larger role in peak reduction \cite{LAVIN2021112484, gundogdu2018battery}. Therefore, deploying the right pricing mechanism is vital for effectively managing a large volume of flexible demand in the future electric grid.

Several electricity pricing schemes are used across the globe, but a select few have gained prominence for their effectiveness in mitigating peak demand. In this work, we focus on two widely implemented approaches: \emph{anytime peak pricing} (AP) and \emph{coincident peak pricing} (CP).  Under AP (used in California and Japan), consumers pay a demand charge based on their individual peak consumption over the billing period \cite{HERTER20072121, KATO201612}. Under CP (used in Texas and the United Kingdom), consumers pay a demand charge based on their consumption during the system-wide peak period \cite{zarnikau2013response, NESO2023Doc130641}.
% While the mechanisms may seem similar, the behavior they induce is fundamentally different: anytime peak pricing encourages flexible users to `flatten' their consumption, whereas coincident peak pricing incentivizes them to `dodge' the aggregate peak.
The key distinction between these two mechanisms lies in the coupling of the consumers' objectives. In AP, a consumer's cost is only a function of their own consumption, whereas in CP, a consumer's cost is a function of the consumption of all consumers on the grid. This coupling presents both opportunities and risks for peak reduction.

The focus of this paper is on understanding which mechanism is better at reducing peak demand. Our main result, Theorem \ref{thm:peaks_baseline}, demonstrates that there is no simple answer to this question. Focusing on the solution concept of approximate Nash equilibrium, a configuration in which each user consumes electricity in an approximately optimal manner given others' behaviors, we show that the answer depends critically on the environmental conditions. More specifically, when the exogenous baseline loads on the system are deterministic and known to all users, CP outperforms AP by enabling coordinated peak avoidance. However, when baseline loads are random and users must rely on probabilistic forecasts, AP outperforms CP because the coordination introduced by CP can backfire, causing users to miscoordinate and enabling the realization of larger peaks. Hence, neither mechanism is universally better; the choice depends on the information structure of the environment.

Since neither dominates, we next investigate whether alternative mechanisms might offer superior performance guarantees in both settings. A key observation is that CP's vulnerability stems from the linearity of its demand charge: consumers pay a per-unit cost for their consumption during the coincident peak period. Under this structure, users minimizing their expected cost effectively minimize the expected peak demand on the system, but this provides no protection against worst-case peak realizations. Motivated by this insight, we move beyond linear pricing structures and instead penalize high-demand scenarios more severely through progressive demand charges, where the demand cost grows superlinearly with consumption. Our results demonstrate that this \emph{progressive peak pricing} mechanism (PP) retains CP's benefits while mitigating its risks: for deterministic baseline loads, PP performs equivalently to CP (Theorem \ref{thm:pp_deterministic}), while for certain classes of random baseline loads, PP outperforms CP (Proposition \ref{prop:progressive_peak_pricing}). However, whether PP maintains this improvement across all probability distributions remains an open question, which we explore through simulation.

Our paper and results contribute to a growing body of literature analyzing equilibrium behavior under electricity pricing mechanisms. This work broadly falls under the category of mechanism design, where the goal is to design pricing structures that incentivize consumer behavior in ways that achieve socially desirable outcomes, such as minimizing peak demand or improving grid efficiency \cite{10384599, 6266724}. Prior work in this area has taken several complementary approaches: analyzing equilibrium outcomes and strategic behavior under various pricing mechanisms \cite{philpott2007coincident, 8667647, baldick2018incentive, 5628271}; characterizing optimal user behavior and demand response strategies in response to peak pricing \cite{liu2013data, ZAKERI201497}; evaluating the real-world efficacy of these mechanisms through field studies and empirical analysis \cite{LAVIN2021112484, EID201615, HERTER20072121}; and developing optimization frameworks for demand-side management under time-varying electricity prices \cite{6266724, muratori2015residential}. Despite this extensive body of work, a direct comparison of AP and CP that identifies the regions of superiority for each mechanism, and how these regions depend on the information structure, has not been adequately explored. Our work addresses this gap by providing both theoretical characterizations of when each mechanism performs best and practical insights into deployment decisions. Moreover, this analysis prompts us to reconsider whether linear demand costs are appropriate for mechanisms aimed at reducing peak load, motivating our introduction of progressive peak pricing.

The remainder of this paper is organized as follows. Section \ref{sec:model} introduces the model of the electricity market, users, and the various pricing mechanisms. We provide our main results and explore our findings in simulation in Section \ref{sec:results}, and we present our conclusions in Section \ref{sec:conclusion}. \textbf{All proofs are available in the Appendix.}

\section{Electricity Market Model}\label{sec:model}

In this paper, we model the strategic interactions between an electricity provider and users, where the provider uses pricing mechanisms to shape user behavior with the goal of reducing peak demand. We model an electricity market as a collection of users who consume electricity over a finite horizon. There are two kinds of users, \emph{flexible} and \emph{inflexible}, defined as follows:

\vspace{.2cm}

\noindent \textbf{\emph{ -- Flexible users:}} We model the flexible users as a set of $N$ \emph{Players}, where Player $i \in [N]$ has a \emph{consumption requirement} of $r_i > 0$ electricity units that must be satisfied over a $T$-period horizon; here, $r_i$ represents flexible demand that can be distributed across periods at the Player's discretion. We denote Player $i$'s consumption in period $t \in [T]$ as $a_i^t \geq 0$, and we refer to their consumption over the full horizon, $a_i = (a_i^1, ..., a_i^T)$, as their \emph{action}. Player $i$'s action space, i.e., the set of feasible consumption profiles, is thus given by
\begin{equation}\label{eq:consumption_constraint}
    \mathcal{A}_i \triangleq \left\{a_i \in \mathbb{R}_{\geq 0}^T \, \Bigg| \, \sum_{t = 1}^T a_i^t \geq r_i \right\}.
\end{equation}
We refer to the actions of all Players, $a = (a_1, ..., a_N)$, as the \emph{action profile}, which belongs to the set $\mathcal{A} = \prod_{i = 1}^N \mathcal{A}_i$.

\vspace{.2cm}

\noindent \textbf{\emph{ -- Inflexible users:}} We model the aggregate behavior of the inflexible users as a profile $b = (b^1, \dots, b^T)$ of \emph{baseline loads} on the electricity grid; here, $b^t$ represents the aggregate electricity consumption requirement of the inflexible users in period $t$, which must be met exactly. We assume that the Players have some forecast of the baseline load; they either know $b$ exactly, or they know the distribution $B$ from which it is sampled. We denote the bounded support of $B$ as $\text{supp}(B)$.

The goal of the electricity provider is to incentivize the flexible users so as to minimize the \emph{peak demand}, i.e., the maximum aggregate load on the grid, which is defined as
\begin{equation}\label{eq:peak_demand}
    P(a;b) \triangleq \max_{t \in [T]} \, \left( b^t + \sum_{i = 1}^N a_i^t \right),
\end{equation}
where $a = (a_1, \dots, a_{N})$ is a given action profile and $b$ is a given baseline load.  
The provider uses a \emph{pricing mechanism}, a system by which users are charged for their electricity consumption, to incentivize the reduction of peak demand. In this work, we study the following two widely implemented pricing mechanisms for peak reduction. 

\vspace{0.2cm}
\noindent \emph{\textbf{-- Anytime peak pricing}} (${\rm AP}$): Player $i$ pays a time-of-use rate $p^t \geq 0$ for the electricity consumed in period $t$ plus a \emph{demand} rate $p^{\rm D} > T \max_t p^t$ for their maximum consumption in \emph{any} period:\footnote{Time-of-use indicates that users pay different rates depending on the time at which they use electricity. Several real-world electricity markets, especially residential ones, impose time-of-use rates without demand rates.}
\begin{equation}\label{eq:ca_pricing}
    J_i^{\rm AP}(a;b) \triangleq \sum_{t = 1}^T \left( p^t a_i^t \right) + p^{\rm D} \max_{t'} a_i^{t'}.
\end{equation}

\begin{figure}
    \centering
    \includegraphics[width=\linewidth]{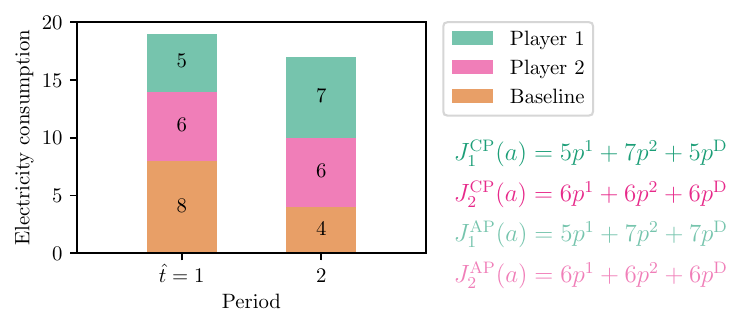}
    \caption{An electricity market with $T = 2$ periods and $N = 2$ Players. The baseline load is $b = (8, 4)$. The Players' consumption requirements and actions are $r_1 = r_2 = 12$ and $a_1 = (5, 7)$,  $a_2 = (6, 6)$, respectively. Under anytime peak pricing, Player $1$ pays a demand charge for their consumption in period $2$, while under coincident peak pricing, they pay a demand charge for their consumption in period $1$.}
    \label{fig:pricing_example}
\end{figure}

\vspace{0.2cm}
\noindent \emph{\textbf{-- Coincident peak pricing}} (${\rm CP}$): Player $i$ pays a time-of-use rate $p^t \geq 0$ for the electricity consumed in period $t$ plus a demand rate $p^{\rm D} > T \max_t p^t$ for their consumption in the coincident peak period:
\begin{equation}\label{eq:tx_pricing}
    J_i^{\rm CP}(a;b) \triangleq \sum_{t = 1}^T \left( p^t a_i^t \right) + p^{\rm D} a_i^{\hat{t}},
\end{equation}
where
\begin{equation}\label{eq:coincident_peak_period}
    \hat{t} \triangleq \argmax_{t} \, b^t + \sum_{i = 1}^N a_i^t
\end{equation}
is the \emph{coincident peak period}\footnote{Note that in the case where the coincident peak is not unique, the demand charge may change depending on which period $\hat{t}$ is chosen. We break ties by choosing the earliest period, but our results are not affected by this choice.}. See Figure \ref{fig:pricing_example} for a comparison of the two mechanisms.

Since $\rm CP$ couples Players' costs through $\hat{t}$, we refer to our market model as an \emph{electricity market game} and denote a specific instance by the tuple $\mathcal{G}^{\rm M} = \left( N, \mathcal{A}, \{ J_i^{\rm M} \}_{i \in [N]}, B \right)$ where $\rm M \in \{AP, CP\}$ denotes the pricing mechanism of the electricity market.  We model the Players' emergent behavior in a given electricity market game as an approximate Nash equilibrium, which is defined as follows:
\begin{definition}
    For an electricity market game $\mathcal{G}^{\rm M}$, an action profile $\bar{a} \in \mathcal{A}$ is an \emph{$\varepsilon$-Nash equilibrium} if
    \begin{equation}\label{eq:e-ne-def}
        \mathbb{E}_B \left[ J_i^{\rm M} (\bar{a}_i, \bar{a}_{-i}; b) \right] \leq \mathbb{E}_B \left[ J_i^{\rm M} (a_i, \bar{a}_{-i}; b) \right] + \varepsilon
    \end{equation}
    for all $a_i \in \mathcal{A}_i, \, i \in [N]$.
\end{definition}
An approximate Nash equilibrium characterizes an outcome $\bar{a} \in \mathcal{A}$ where each Player's action is approximately optimal given everyone else's behavior. The following Proposition ensures that an $\varepsilon$-Nash equilibrium is guaranteed to exist in any electricity market game for any $\varepsilon > 0$.
\begin{proposition}\label{prop:eq_exists}
    For any electricity market game $\mathcal{G}^{\rm M}$ and $\varepsilon > 0$, the set
    \begin{equation*}
        \mathcal{A}^{\rm M}(\varepsilon) \triangleq \{\bar{a} \in \mathcal{A} \, | \, \text{$\bar{a}$ is an $\varepsilon$-Nash equilibrium} \}
    \end{equation*}
    is nonempty for $\rm M \in \{AP, CP \}$.
\end{proposition}
Proposition \ref{prop:eq_exists} ensures that our model produces well-defined outcomes in general scenarios. With existence guaranteed, we now turn to evaluating the quality of these equilibria. Our metric for comparison is the peak demand induced at equilibrium, which we define as the worst-case peak across all approximate equilibria:
\begin{definition}
    For an electricity market game $\mathcal{G}^{\rm M}$, the \emph{equilibrium peak demand} is
    \begin{equation}\label{eq:worst_case_peak_definition}
        \mathcal{P}\left( \mathcal{G}^{\rm M} \right) \triangleq \lim_{\varepsilon \to 0^+} \; \sup_{\bar{a} \in \mathcal{A}^{\rm M} (\varepsilon)} \;
        \sup_{b \in \text{supp}(B)}
        \, P(\bar{a}; b).
    \end{equation}
\end{definition}
Observe that we consider the case when $\varepsilon \to 0^+$ to focus on behavior arbitrarily close to exact equilibrium, avoiding artifacts from suboptimal behavior; in the Appendix, we show that this limit is in fact well-defined. In the following section, we use this metric to compare anytime and coincident peak pricing.

% Having formally defined our model of an electricity market, we can now describe more formally the questions we address in this work. First, what does the Players' equilibrium behavior look like under each of the pricing mechanisms? Second, how does the peak demand vary across pricing mechanisms in the presence of deterministic and random exogenous baseline loads? And finally, from this analysis, can we conclude which mechanism one should use? We answer each of these questions in the following section.

\section{Peak Analysis}\label{sec:results}

This section contains our main results on peak demand under coincident and anytime peak pricing. First, we compare  anytime and coincident peak pricing under deterministic and random baseline loads. Then, we present a third mechanism as an alternative to anytime and coincident peak pricing and characterize its performance analytically and in simulation.

\subsection{Anytime and Coincident Peak Pricing}\label{sec:results_no_baseline}

Our main result compares the size of the peak under anytime and coincident peak pricing in different kinds of markets.

\begin{theorem}\label{thm:peaks_baseline}
    Consider an electricity market game $\mathcal{G}^{\rm M}$ with consumption requirements $r_1, \dots, r_N$, time-of-use rates $p^1, \dots, p^T$, and demand rate $p^{\rm D}$.
    \begin{enumerate}[label=(\alph*)]
        \item If the baseline load $b$ is deterministic, then the equilibrium peak demand under coincident peak pricing is no greater than that under anytime peak pricing:
        \begin{equation}\label{eq:baseline_relation}
            \mathcal{P} \left( \mathcal{G}^{\rm CP} \right)
            \leq
            \mathcal{P} \left( \mathcal{G}^{\rm AP} \right).
        \end{equation}

        \item If the baseline load $b$ is random, then \eqref{eq:baseline_relation} may not hold; in particular, the equilibrium peak demand under anytime peak pricing may be less than, greater than, or equal to that under coincident peak pricing.
    \end{enumerate}
\end{theorem}

Theorem \ref{thm:peaks_baseline} establishes that if the baseline loads are perfectly forecast, then coincident peak pricing is always better, but if not, then this may not be the case. This reversal occurs because under uncertainty, CP incentivizes Players to minimize expected cost, which can lead to concentration of flexible load in periods deemed unlikely to be peaks. When these forecasts are wrong, this concentration exacerbates peak demand. To make this explicit, we present and compare the closed-form expressions (derived in the Appendix) for the peak demands in two special settings.

\vspace{0.1cm}
\noindent \textbf{Example 1: Deterministic Baseline} If the baseline load $b$ is deterministic, then the equilibrium peak demands are
\begin{align}
    \mathcal{P} \left( \mathcal{G}^{\rm CP} \right) &= \max \left( \max_{t \in [T]} b^t, \, \sum_{t = 1}^T \frac{b^t}{T} + \sum_{i = 1}^N \frac{r_i}{T} \right), \label{eq:tx_peak_baseline} \\
    \mathcal{P} \left( \mathcal{G}^{\rm AP} \right) &= \max_{t \in [T]} b^t + \sum_{i = 1}^N \frac{r_i}{T}. \label{eq:ca_peak_baseline}
\end{align}
For example, in a market with $N = 2$ Players, $T = 2$ periods, consumption requirements $r_1 = r_2 = 6$, and baseline load $b = (12, 0)$, we have $\mathcal{P} \left( \mathcal{G}^{\rm CP} \right) = 12$ and $\mathcal{P} \left( \mathcal{G}^{\rm AP} \right) = 18$, so the peak is $50$\% larger under $\rm AP$.

\vspace{0.1cm}
\noindent \textbf{Example 2: Random Baseline} If the baseline load $b \sim B$ is drawn from a distribution satisfying
\begin{gather}
    \max \left( \text{supp} \left( B^t \right) \right) = \bar{b} \; \; \forall \, t \in [T], \label{eq:equal_support} \\
    \mathbb{P} \left[ \hat{t} = t \right] = \mathbb{P}\left[b^t = \max_{s \in [T]} b^s \right] \; \forall \, t \in [T], \, a \in \mathcal{A}, \label{eq:action_doesnt_matter}
\end{gather}
then the equilibrium peak demands are
\begin{align}
    \mathcal{P} \left( \mathcal{G}^{\rm CP} \right) &= \bar{b} + \sum_{i = 1}^N r_i, \\
    \mathcal{P} \left( \mathcal{G}^{\rm AP} \right) &= \bar{b} + \sum_{i = 1}^N \frac{r_i}{T}.
\end{align}
For example, in a market with $N = 2$ Players, $T = 2$ periods, consumption requirements $r_1 = r_2 = 6$, and baseline load
\begin{equation*}
    b = \begin{cases}
        (12, 0) & \text{with probability 0.6} \\
        (0, 12) & \text{with probability 0.4},
    \end{cases}
\end{equation*}
we have $\mathcal{P} \left( \mathcal{G}^{\rm CP} \right) = 24$ and $\mathcal{P} \left( \mathcal{G}^{\rm AP} \right) = 18$, so the peak is $25$\% smaller under $\rm AP$.

\begin{figure}
    \centering
    \includegraphics[width=\linewidth]{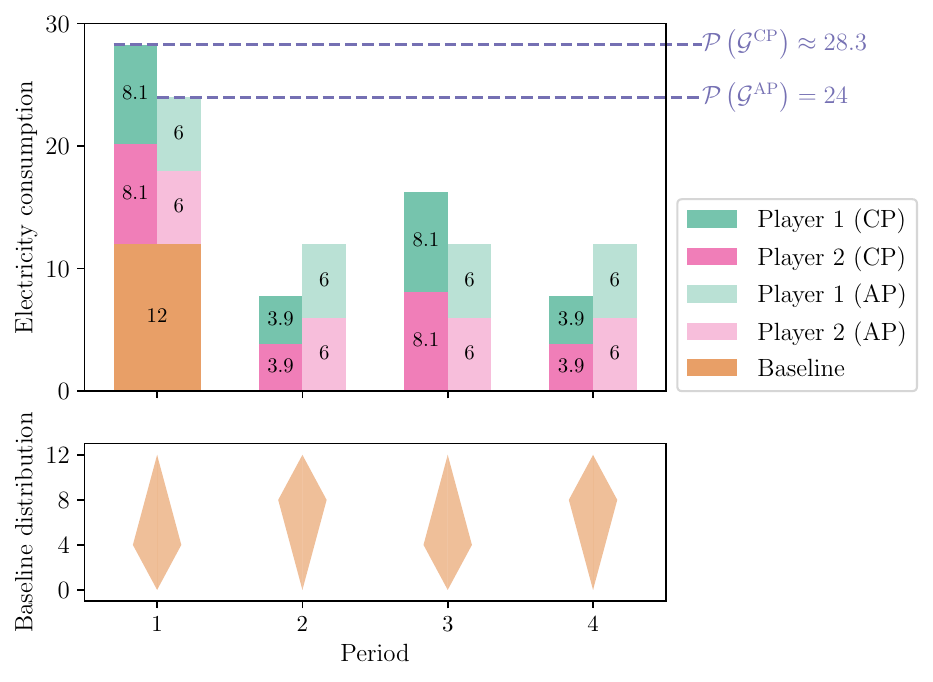}
    \caption{An electricity market with $T = 4$ periods and $N = 2$ Players with consumption requirements $r_1 = r_2 = 24$. The bottom plot shows the baseline load $b^t$, which is independently drawn from a scaled triangular distribution with mode $4$ if $t$ is odd or $8$ if $t$ is even. The top plot shows the Players' equilibrium actions for coincident peak (darker shade) and anytime peak (lighter shade) pricing, along with a worst-case baseline realization (dark orange); observe that the peak is larger under coincident peak pricing.}
    \label{fig:random_example}
\end{figure}

\vspace{0.1cm}
These examples demonstrate that neither mechanism universally dominates: $\rm CP$ excels when baseline loads are known, while $\rm AP$ can outperform when they are uncertain. In general, it is difficult to characterize the equilibrium peak demands for arbitrary baseline load distributions, but in simulation, we observe that coincident peak pricing can induce larger peaks than anytime peak pricing in several scenarios. Figure \ref{fig:random_example} shows a scenario with continuous baseline distributions (unlike the discrete distribution in Example 2), demonstrating that CP's disadvantage extends beyond these specific cases.

To decide which mechanism should be used in the presence of random baseline loads, one can assume that the baseline load is sampled from some distribution and determine whether anytime or coincident peak pricing performs better in that specific setting. Alternatively, one can ask whether there exists a different mechanism that retains the desirable features of both anytime and coincident peak pricing for general distributions. Such a mechanism would perform at least as well as coincident peak pricing in the deterministic setting while outperforming it in random settings. In the following section, we pursue this direction, proposing and evaluating a third mechanism for peak reduction.

\subsection{Progressive Peak Pricing}

In this section, we study an alternative to coincident and anytime peak pricing and compare it with the two. We refer to this mechanism as progressive peak pricing:

\vspace{0.2cm}
\noindent \emph{\textbf{-- Progressive peak pricing}} (${\rm PP}$): Player $i$ pays a time-of-use rate $p^t \geq 0$ for the electricity consumed in period $t$ plus a demand rate $p^{\rm D}$ for their consumption during the coincident peak period raised to a power $\delta \geq 1$:
\begin{equation}
    J_i^{\rm PP}(a; b) \triangleq 
    \sum_{t = 1}^T \left( p^t a_i^t \right) + p^{\rm D} \left( a_i^{\hat{t}}\right)^\delta,
\end{equation}
where $\hat{t}$ satisfies \eqref{eq:coincident_peak_period}. Progressive peak pricing is a generalization of coincident peak pricing in which Players pay superlinear demand charges for their consumption in the coincident peak period. Observe that when $\delta = 1$, progressive peak pricing is equivalent to coincident peak pricing.

Our first result on progressive peak pricing establishes that it preserves the advantages of coincident peak pricing in the deterministic setting:
\begin{theorem}\label{thm:pp_deterministic}
    Consider an electricity market game $\mathcal{G}^{\rm M}$ with consumption requirements $r_1, \dots, r_N$, time-of-use rates $p^1 = \dots = p^T$,\footnote{The conference version of this paper \cite{shah2026playing} states Theorem~\ref{thm:pp_deterministic} for arbitrary time-of-use rates $p^1, \dots, p^T$, which is incorrect when $\delta > 1$. For example, with $N = 1$, $T = 2$, $b = (10, 0)$, $r_1 = 1$, $p^1 = 0$, $p^2 = 1$, $p^{\rm D} = 3$, and $\delta = 2$, the Player's unique optimal action under progressive peak pricing is $a_1 = (1/6, 5/6)$, yielding a peak of $10 + 1/6$, whereas the peak under coincident peak pricing is $10$.} and demand rate $p^{\rm D}$. If the baseline load $b$ is deterministic, then for any $\delta \geq 1$, the equilibrium peak demand under progressive peak pricing is equal to that under coincident peak pricing:
    \begin{equation}\label{eq:pp_baseline_relation}
        \mathcal{P} \left( \mathcal{G}^{\rm CP} \right)
        =
        \mathcal{P} \left( \mathcal{G}^{\rm PP} \right)
        \leq
        \mathcal{P} \left( \mathcal{G}^{\rm AP} \right).
    \end{equation}
\end{theorem}

\begin{figure*}
    \centering
    \includegraphics[width=\linewidth]{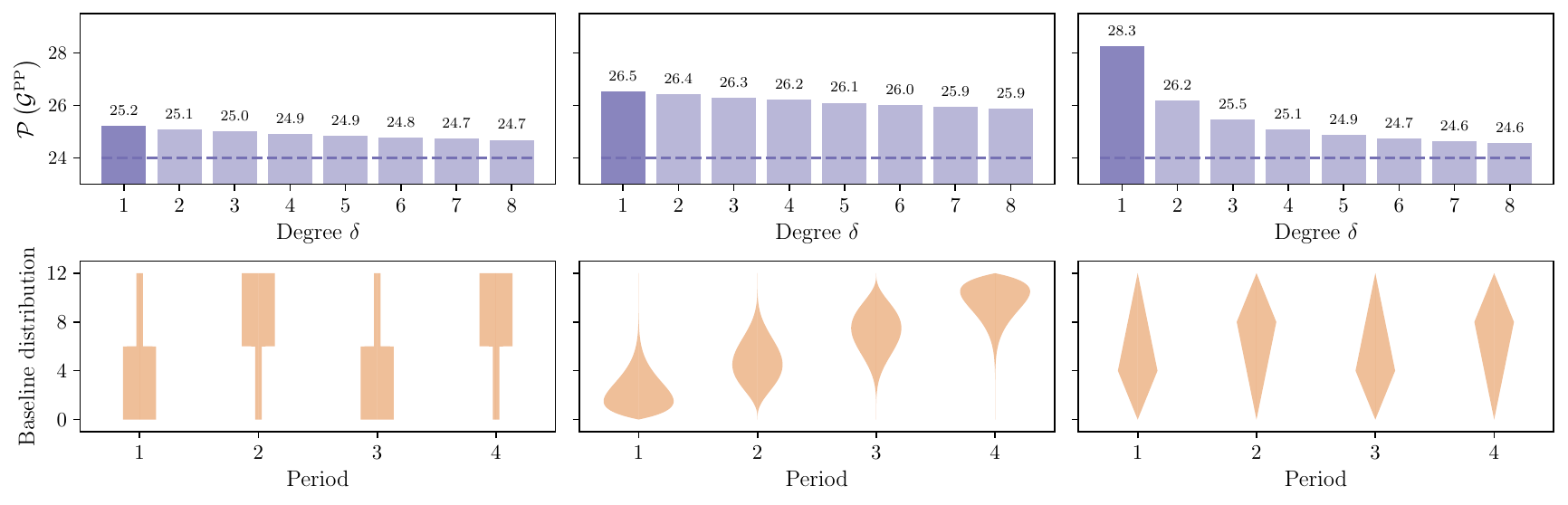}
    \caption{Comparisons of anytime, coincident, and progressive peak pricing in an electricity market with $T = 4$ periods and $N = 2$ Players with consumption requirements $r_1 = r_2 = 24$. We consider three different baseline load distributions: conditional uniform (left), Beta (center), and triangular (right). The equilibrium peak demand (lavender) is shown for each distribution under progressive peak pricing for various values of $\delta$ (top row); observe that as $\delta$ increases, the peak decreases. The equilibrium peak demand under coincident peak pricing is the first bar in each plot (darker shade), and the equilibrium peak demand under anytime peak pricing is $24$ for all three scenarios (dashed purple line).}
    \label{fig:sims}
\end{figure*}

Theorem \ref{thm:pp_deterministic} establishes that progressive peak pricing matches coincident peak pricing's performance in the deterministic setting, meaning that progressive demand charges cannot induce larger peaks when baseline loads are known and time-of-use rates are uniform.

We now examine whether it offers advantages under uncertainty. Figure \ref{fig:sims} compares anytime, coincident, and progressive peak pricing in three examples with baseline loads sampled from different families of distributions. As shown in the plot, in all three scenarios, progressive peak pricing outperforms coincident peak pricing for all choices of $\delta > 1$. Furthermore, as $\delta$ increases, the peaks decrease, indicating that progressive peak pricing can bridge the gap between anytime and coincident peak pricing.

In several simulations, we have observed that $\rm PP$ is always at least as good as $\rm CP$. While characterizing progressive peak pricing's performance for arbitrary distributions remains open, we can establish this result analytically for the family of distributions introduced in Example 2:

\begin{proposition}\label{prop:progressive_peak_pricing}
    Consider an electricity market game $\mathcal{G}^{\rm M}$ with consumption requirements $r_1, \dots, r_N$, time-of-use rates $p^1 = \dots = p^T$, and demand rate $p^{\rm D}$. If the baseline load $b$ is  drawn from a continuous distribution $B$ satisfying \eqref{eq:equal_support} and \eqref{eq:action_doesnt_matter}, then for any $\delta > 1$, the equilibrium peak demand under progressive peak pricing is no greater than that under coincident peak pricing:
    \begin{equation}\label{eq:pp_random_baseline_relation}
        \mathcal{P} \left( \mathcal{G}^{\rm CP} \right)
        \geq
        \mathcal{P} \left( \mathcal{G}^{\rm PP} \right)
        \geq
        \mathcal{P} \left( \mathcal{G}^{\rm AP} \right).
    \end{equation}
\end{proposition}

Proposition \ref{prop:progressive_peak_pricing} establishes that for distributions satisfying \eqref{eq:equal_support} and \eqref{eq:action_doesnt_matter}, progressive peak pricing provides robust performance: it matches $\rm CP$ when baseline loads are known and improves upon it when they are uncertain, making it a clear choice for these settings. We conjecture that this result holds for much broader families of distributions, which we aim to show in future work.

\section{Conclusion}\label{sec:conclusion}

In this work, we compare anytime and coincident peak pricing with respect to their abilities to mitigate peak demand in electricity markets with flexible and inflexible users. We first establish a connection between electricity markets and game theory and show that in our model, an $\varepsilon$-Nash equilibrium exists for either mechanism. Then, in our main results, we show that with fixed baseline loads on the grid, coincident peak pricing always performs at least as well as anytime peak pricing, but with random baseline loads, this need not be true. We propose progressive peak pricing as an alternative and show that it performs well in deterministic and specialized random settings, offering a robust solution across grid conditions. Simulation studies corroborate and generalize our findings to other families of distributions.

This work opens several new directions for future research. First, our analytical results for progressive peak pricing under uncertainty apply to a limited family of distributions. Extending these results to broader distribution classes is an important direction for future work. Second, the performance of more intuitive progressive demand structures, such as convex piecewise linear functions, should be studied. Finally, augmenting our model to include more realistic real-world phenomena, such as network constraints and losses, could extend the scope and applicability of these results.

\bibliographystyle{plain}
\bibliography{references}

\appendix
\newcounter{mycounter}
\setcounter{mycounter}{1}
\newtheorem{claim}{Claim}[mycounter]

To simplify notation, we assume $p^1 \leq \dots \leq p^T$ in all proofs (i.e., the first period has the cheapest time-of-use rate); this is without loss of generality. Furthermore, we omit writing $b$ in $J^{\rm M}$ when the dependence is clear.

\subsection{Proofs of Proposition \ref{prop:eq_exists}, Theorem \ref{thm:peaks_baseline}, and Theorem \ref{thm:pp_deterministic}}\label{sec:proof_peaks_baseline}

In this section, we prove Theorems \ref{thm:peaks_baseline} and \ref{thm:pp_deterministic}; on the way, we prove Proposition \ref{prop:eq_exists}. Note that Theorem \ref{thm:peaks_baseline}(b) follows from Examples 1 and 2 in the main text (which in turn follow from Theorem \ref{thm:peaks_baseline}(a) and Proposition \ref{prop:progressive_peak_pricing}). Thus, we need only prove Theorem \ref{thm:peaks_baseline}(a) by establishing \eqref{eq:tx_peak_baseline} and \eqref{eq:ca_peak_baseline}.

Let $b^{t_{\max}} = \max_{t \in [T]} b^t$ (not to be confused with the maximum of the support $\bar{b}$ in \eqref{eq:equal_support}). Assuming that \eqref{eq:tx_peak_baseline} and \eqref{eq:ca_peak_baseline} are correct, we first establish \eqref{eq:baseline_relation}. If $\mathcal{P} \left( \mathcal{G}^{\rm CP} \right) = b^{t_{\max}}$, then \eqref{eq:baseline_relation} is clearly satisfied. Otherwise, we have
\begin{align*}
    \mathcal{P} \left( \mathcal{G}^{\rm CP} \right) &= \frac{\sum_{t = 1}^T b^t + \sum_{i = 1}^N r_i}{T} \\
    &\leq \frac{T b^{t_{\max}} + \sum_{i = 1}^N r_i}{T} = \mathcal{P} \left( \mathcal{G}^{\rm AP} \right).
\end{align*}

We now establish \eqref{eq:tx_peak_baseline} and \eqref{eq:ca_peak_baseline}.

\subsubsection{Anytime peak pricing (${\rm AP}$)} First, notice that if we impose the constraint $\max_t a_i^t = m$ on Player $i$'s allocation, then the unique optimal solution is a water-filling one that allocates up to $m$ in increasing periods, i.e.,
\begin{equation*}
    a_i^t = [\min( m, r_i - m(t - 1))]_+.
\end{equation*}
This solution incurs cost
\begin{equation}\label{eq:water_filling_solution}
     J_i^{\rm AP}(a_i, a_{-i}) = m \left( p^{\rm D} + \sum_{s = 1}^{t'} p^s \right) + p^{t' + 1}(r_i - m t')
\end{equation}
where $t' = \left\lfloor \frac{r_i}{m} \right\rfloor$. Thus, Player $i$'s optimization problem reduces to that of finding the value of $m$ that minimizes \eqref{eq:water_filling_solution}.

As an intermediate step towards this goal, we analyze the function
\begin{equation}\label{eq:optimal_spreading}
    h(t) = \frac{1}{t} \left( p^{\rm D} + \sum_{s = 1}^t p^s \right)
\end{equation}
defined on the integer domain $t \in [T]$.

\begin{claim}
    The function $h$ is unimodal in $t$, i.e., it is monotonically nonincreasing for $t < t^*$ and monotonically nondecreasing for $t > t^*$ for some $t^* \in [T]$.
\end{claim}

\begin{proof}
    Showing that $h$ is unimodal is equivalent to showing that for all $t \geq 1$, $h(t) \leq h(t + 1) \implies h(t + 1) \leq h(t + 2)$. We have
    \begin{align}
        &h(t) \leq h(t + 1) \nonumber \\
        \iff &\frac{p^{\rm D} + \sum_{s=1}^t p^s}{t} \leq \frac{p^{\rm D} + \sum_{s=1}^{t + 1} p^s}{t + 1} \nonumber \\
        \iff &(t + 1) \left( p^{\rm D} + \sum_{s=1}^t p^s \right) \leq t \left( p^{\rm D} + \sum_{s=1}^{t + 1} p^s \right) \nonumber \\
        \iff  &p^{\rm D} + \sum_{s=1}^t p^s \leq t p^{t+1} \label{eq:better_to_shift} \\
        % \iff  &p^{\rm D} + \sum_{s=1}^{t + 1} p^s \leq (t + 1) p^{t+1} \nonumber \\
        \implies  &p^{\rm D} + \sum_{s=1}^{t + 1} p^s \leq (t + 1) p^{t+2} \nonumber \\
        \iff &(t + 2) \left( p^{\rm D} + \sum_{s=1}^{t + 1} p^s \right) \leq (t + 1) \left( p^{\rm D} + \sum_{s=1}^{t + 2} p^s \right) \nonumber \\
        \iff &h(t + 1) \leq h(t + 2), \nonumber
    \end{align}
    so $h$ is unimodal. 
\end{proof}

We use this property to characterize the optimal value $m$. 

\begin{claim}\label{claim:divides_evenly}
    There exists a real number $m$ that minimizes \eqref{eq:water_filling_solution} and divides $r_i$.
\end{claim}

\begin{proof}
Let $t^*$ denote the smallest minimizer of the function $h$. First, we analyze the cost of the water-filling solution for any value of $m$ for which $t' = \left\lfloor \frac{r_i}{m} \right\rfloor > t^*$. Suppose that $\left\lfloor \frac{r_i}{m} \right\rfloor \neq \frac{r_i}{m}$, meaning that $a_i^{t' + 1} = r_i - m t' > 0$. Since $h$ is unimodal, we know that $h(t') \leq h(t' + 1)$ which implies \eqref{eq:better_to_shift}. Substituting $t = t'$ in \eqref{eq:better_to_shift} and multiplying both sides by $a_i^{t' + 1}$ yields
\begin{equation}\label{eq:better_to_shift_2}
    \left(p^{\rm D} + \sum_{s=1}^{t'} p^s \right)\frac{a_i^{t' + 1}}{t'} \leq p^{t'+1} a_i^{t' + 1},
\end{equation}
meaning that removing all of the consumption from period $t' + 1$ and splitting it evenly between periods $1, \dots, t'$ (yielding another water-filling solution with $m = r_i / t'$) does not increase costs. The opposite argument (shifting consumption from periods $1, \dots, t'$ to $t' + 1$ does not increase costs) follows symmetrically for $t' < t^*$.
\end{proof}

We can now establish \eqref{eq:ca_peak_baseline}.

\begin{claim}
    If the baseline load $b$ is deterministic, then
\begin{equation*}
    \mathcal{P} \left( \mathcal{G}^{\rm AP} \right) = \max_{t \in [T]} b^t + \sum_{i = 1}^N \frac{r_i}{T}.
\end{equation*}
\end{claim}

\begin{proof}
Claim \ref{claim:divides_evenly} implies that restricting attention to values of $m$ that divide $r_i$ is sufficient, in which case \eqref{eq:water_filling_solution} reduces to $m h(t)$. Hence, it is optimal for Player $i$ to allocate their consumption evenly across the first $t$ periods, where $t$ is any element of $\mathcal{T}$, the set of minimizers of $h$. The smallest minimizer of $h$, $t^*$, yields $a_i^t = \frac{r_i}{t^*}$ for all $t \leq t^*$, which is the optimal action for Player $i$ that results in their largest possible contribution to the peak demand. Thus, if every Player $i \in [N]$ allocates $\frac{r_i}{t^*}$ to the first $t^*$ periods, then the load on the grid in any period $t \leq t^*$ is $\frac{1}{t^*} \sum_i r_i$. However, under our assumption that $p^{\rm D} > T \max_t p^t$, we have that $t^* = T$ is the unique minimizer of $h$. Thus, each Player allocates their consumption requirement evenly over all $T$ periods; taking the maximum of the baseline load $b$ over $t$ yields \eqref{eq:ca_peak_baseline}.
\end{proof}

Here, note that we have characterized each Player's optimal action, so we have also established Proposition \ref{prop:eq_exists} for anytime peak pricing.

\subsubsection{Coincident ($\rm CP$) and progressive peak pricing ($\rm PP$)} Note that when the baseline load is deterministic and the time-of-use rates are equal, the exponent $\delta$ does not change the equilibrium peak demand; hence, we can analyze coincident and progressive peak pricing as one. Throughout, the time-of-use rates $p^1, \dots, p^T$ are arbitrary for coincident peak pricing (as in Theorem~\ref{thm:peaks_baseline}), whereas for progressive peak pricing with $\delta > 1$ we assume $p^1 = \dots = p^T$ (as in Theorem~\ref{thm:pp_deterministic}); where the arguments for the two mechanisms differ, we treat them separately.

The proof proceeds by showing that
\begin{enumerate}[label=(\alph*)]
    \item for any $\varepsilon > 0$, the set of $\varepsilon$-Nash equilibria $\mathcal{A}^{\rm CP}(\varepsilon)$ is nonempty
    \item the equilibrium peak demand $\mathcal{P} \left( \mathcal{G}^{\rm CP} \right)$ is as written in \eqref{eq:tx_peak_baseline}.
\end{enumerate}
First, notice that
\begin{gather*}
    b^{t_{\max}} > \frac{\sum_{t = 1}^T b^t + \sum_{i = 1}^N r_i}{T} \\
    \iff 
    \sum_{t = 1}^T (b^{t_{\max}} - b^t) - \sum_{i = 1}^N r_i \triangleq \Delta > 0.
\end{gather*}
Thus, we split our analysis into two cases: $\Delta > 0$ and $\Delta \leq 0$. For each case, we address (a) and (b). 

\stepcounter{mycounter}
\begin{claim}
    If $\Delta > 0$, then for any $\varepsilon > 0$, the set of $\varepsilon$-Nash equilibria $\mathcal{A}^{\rm CP}(\varepsilon)$ ($\mathcal{A}^{\rm PP}(\varepsilon)$) is nonempty.
\end{claim}

\begin{proof}
If $\Delta > 0$, then it is possible for the Players to allocate their consumption such that no Player faces any demand charges. Let $\mathcal{T}_{\max}$ denote the set of periods in which $b^t = b^{t_{\max}}$, and let $\{\tau_1, \dots, \tau_k\} = [T] \setminus \mathcal{T}_{\max}$. For any
% 
% $\varepsilon \in (0, T^2 p^T \Delta)$,
% 
sufficiently small $\varepsilon$, let $\eta = \frac{\varepsilon}{T p^T}$, and consider the action profile defined by the following procedure:
\begin{itemize}
    \item If $r_{1} < b^{t_{\max}} - b^{\tau_1} - \eta$, then $\bar{a}_1^{\tau_1} = r_1$. Otherwise, take $\bar{a}_1^{\tau_1} = b^{t_{\max}} - b^{\tau_1} - \eta$ and proceed to $\tau_2$.  Proceed in this fashion until all of Player $1$'s consumption requirement has been allocated (say, in period $\tau_j$).
    \item If $r_{2} < b^{t_{\max}} - b^{\tau_j} - \bar{a}_1^{\tau_j} - \eta$, then $\bar{a}_2^{\tau_j} = r_2$. Otherwise, take $\bar{a}_2^{\tau_j} = b^{t_{\max}} - b^{\tau_j} - \bar{a}_1^{\tau_j} - \eta$ and proceed to $\tau_{j + 1}$. Proceed in this fashion until all Players' consumption requirements have been allocated.
\end{itemize}

% define the action profile
% \begin{equation*}
%     \bar{a}_i^t =  \frac{r_i}{\sum_{i = 1}^N r_i} [b^{t_{\max}} - b^t - \eta]_+.
% \end{equation*}

% \begin{equation*}
%     \bar{a}_i^t = \max \left( 0, \min \left( \left( b^{t_{\max}} - b^t - \eta \right) \frac{r_i}{\sum_{i = 1}^N r_i}, r_i - \sum_{s = 1}^{t - 1} \bar{a}_i^s \right) \right).
% \end{equation*}

Essentially, the Players sequentially allocate their consumption in a water-filling manner such that the total consumption is at most $b^{t_{\max}} - \eta$ in timesteps $t \notin \mathcal{T}_{\max}$. Note that under this profile, each Player incurs a demand charge of $0$, since none of them consume any electricity in any periods $t \in \mathcal{T}_{\max}$. We will show that $\bar{a}$ is an $\varepsilon$-Nash equilibrium. Consider the optimization problem Player $i$ seeks to solve given $\bar{a}_{-i}$. It is readily verified that in the case with zero baseline load, the optimal solution for a given peak value (in this case, $b^{t_{\max}}$) is a water-filling one that allocates up to the peak value starting from the cheapest period, i.e.,
\begin{equation*}
    \underline{a}_i^t = \max \left( 0, \min \left( b^{t_{\max}} - b^t - \sum_{j \neq i} \bar{a}_j^t, r_i - \sum_{s = 1}^{t - 1} \underline{a}_i^s \right) \right).
\end{equation*}
This solution would be optimal for the case with nonzero baseline load if the tie-breaking rule were such that $\hat{t} \in \mathcal{T}_{\max}$, since $\underline{a}_i^t = 0$ for all $t \in \mathcal{T}_{\max}$. Assuming this to be the case, we have that
\begin{equation}\label{eq:tx_e_eq_lb}
    J_i^{\rm CP}(a_i, \bar{a}_{-i}) \geq J_i^{\rm CP}(\underline{a}_i, \bar{a}_{-i}) \quad \forall \, a_i \in \mathcal{A}_i.
\end{equation}
However, we also have that the difference in costs $J_i^{\rm CP}(\bar{a}_i, \bar{a}_{-i}) - J_i^{\rm CP}(\underline{a}_i, \bar{a}_{-i})$ is at most
\begin{equation*}
     \sum_{t = 1}^T p^t (\bar{a}_i^t - \underline{a}_i^t) \leq \sum_{t = 1}^T p^t \eta \leq Tp^T \eta = \varepsilon,
\end{equation*}
meaning that even for a favorable tie-breaking rule that yields no demand charges for Player $i$, the savings are at most $\varepsilon$. Since $\underline{a}_i$ is a lower bound on the cost of Player $i$'s optimal action given $\bar{a}_{-i}$ \eqref{eq:tx_e_eq_lb}, we have that $\bar{a}$ is an $\varepsilon$-Nash equilibrium and (a) is verified. For progressive peak pricing with $p^1 = \dots = p^T$, the argument is even simpler: under $\bar{a}$, each Player $i$ pays no demand charge and a time-of-use cost of $p^1 r_i$, which is the minimum cost of any action in $\mathcal{A}_i$, so $\bar{a}$ is in fact an exact Nash equilibrium.
\end{proof}

\begin{claim}\label{claim:delta_pos_peak}
    If $\Delta > 0$, then $\mathcal{P} \left( \mathcal{G}^{\rm CP} \right) = \mathcal{P} \left( \mathcal{G}^{\rm PP} \right) = b^{t_{\max}}.$
\end{claim}

\begin{proof}
Observe that from the previous definition of $\bar{a}$, we have that $\mathcal{P} \left( \mathcal{G}^{\rm CP} \right) \geq b^{t_{\max}}$. We will now show that $b^{t_{\max}}$ is also an upper bound for $\mathcal{P} \left( \mathcal{G}^{\rm CP} \right)$. For any sufficiently small $\varepsilon > 0$, let
\begin{equation}\label{eq:eta_upper_bound}
    \eta = \begin{cases}
        \frac{2 NT \varepsilon}{p^{\rm D} - p^T} & \text{under coincident peak pricing,} \\
        NT \left( \frac{2 \varepsilon}{p^{\rm D}} \right)^{1/\delta} & \text{under progressive peak pricing,}
    \end{cases}
\end{equation}
and let $\breve{a}$ be an action profile for which $b^{\breve{t}} + \sum_{i = 1}^N \breve{a}_i^{\breve{t}} - b^{t_{\max}} = \eta$ for some $\breve{t}$. Then, there must be at least one period $s$ where $b^s + \sum_{i = 1}^N \breve{a}_i^s + \frac{\eta}{T} < b^{t_{\max}}$. Let Player $i$ be any Player who consumes at least $\frac{\eta}{N}$ units of electricity in period $\breve{t}$. They can shift $\frac{\eta}{NT}$ units of this consumption to period $s$, which yields savings
\begin{equation*}
    p^{\rm D} \left( \left( \breve{a}_i^{\breve{t}} \right)^\delta - \left( \breve{a}_i^{\breve{t}} - \frac{\eta}{NT} \right)^\delta \right) + \left( p^{\breve{t}} - p^s \right) \frac{\eta}{NT}.
\end{equation*}
Under coincident peak pricing ($\delta = 1$), these savings are at least $\left( p^{\rm D} - p^T \right) \frac{\eta}{NT} = 2 \varepsilon > \varepsilon$. Under progressive peak pricing with $p^1 = \dots = p^T$, the second term vanishes; moreover, since $z \mapsto z^\delta$ is convex on $\mathbb{R}_{\geq 0}$ with $0^\delta = 0$, it is superadditive, so the savings are at least $p^{\rm D} \left( \frac{\eta}{NT} \right)^\delta = 2 \varepsilon > \varepsilon$. Thus, for fixed $\varepsilon$, we have $b^{t_{\max}} + \eta$ as an upper bound on the size of the coincident peak. Taking the limit as $\varepsilon \to 0^+$ yields $\mathcal{P} \left( \mathcal{G}^{\rm CP} \right) \leq b^{t_{\max}}$, so (b) is verified.
\end{proof}

Next, we consider the case where $\Delta \leq 0$, meaning that at least one Player must face demand charges. The proof in this case follows similarly. First, we establish (a).

\begin{claim}
    If $\Delta \leq 0$, then for any $\varepsilon > 0$, the set of $\varepsilon$-Nash equilibria $\mathcal{A}^{\rm CP}(\varepsilon)$ ($\mathcal{A}^{\rm PP}(\varepsilon)$) is nonempty.
\end{claim}

\begin{proof}
Let $t_{\max}$ be the first period in which $b^t = b^{t_{\max}}$. Consider the action profile
\begin{equation*}
    \bar{a}_i^t = \begin{cases}
        \frac{r_i}{\sum_{j = 1}^N r_j} \left(b^{t_{\max}} - b^t - \frac{\Delta + \eta}{T} + \eta \right) & t = t_{\max} \\
        \frac{r_i}{\sum_{j = 1}^N r_j} \left(b^{t_{\max}} - b^t - \frac{\Delta + \eta}{T} \right) & t \neq t_{\max},
    \end{cases}
\end{equation*}
where $\eta = \frac{\varepsilon}{2 (\delta p^{\rm D} \bar{r}^{\delta - 1} + p^T)}$, $\bar{r} = \max_{j} r_j$, and $\varepsilon$ is sufficiently small. It is readily verified that $\bar{a} \in \mathcal{A}$ and that the action profile is such that the total load on the grid is equal in all periods except period $t_{\max}$, which has an additional $\eta$ units of consumption. We will show that $\bar{a}$ is an $\varepsilon$-Nash equilibrium. Player $i$'s optimal action $\underline{a}_i$ in response to $\bar{a}_{-i}$, assuming a favorable tie-breaking rule that selects $\hat{t} = t_{\max}$, is a water-filling one that allocates load up to $b^{t_{\max}} + \sum_{j \neq i} \bar{a}_j^{t_{\max}}$ starting from the cheapest period. However, the difference in costs $J_i^{\rm CP}(\bar{a}_i, \bar{a}_{-i}) - J_i^{\rm CP}(\underline{a}_i, \bar{a}_{-i})$ is upper bounded by
\begin{align*}
    &\mathrel{\phantom{=}} p^{\rm D} \left( \left( \bar{a}_i^{t_{\max}} \right)^\delta - \left[ \bar{a}_i^{t_{\max}} - \frac{r_i}{\sum_{j = 1}^N r_j} \eta \right]_+^\delta \right) \\
    &\mathrel{\phantom{=}} + \left( p^{t_{\max}} - p^1 \right) \frac{r_i}{\sum_{j = 1}^N r_j} \eta \\
    &\leq \delta p^{\rm D} \left( \bar{a}_i^{t_{\max}} \right)^{\delta - 1} \eta + p^T \eta \\
    &\leq \left( \delta p^{\rm D} \bar{r}^{\delta - 1} + p^T \right) \eta < \varepsilon,
\end{align*}
where the first inequality uses the mean value theorem and $r_i / \sum_{j = 1}^N r_j \leq 1$, and the second uses $\bar{a}_i^{t_{\max}} \leq r_i \leq \bar{r}$. (Under progressive peak pricing with $p^1 = \dots = p^T$, the time-of-use term is zero.) Hence, $\bar{a}$ is an $\varepsilon$-Nash equilibrium and (a) is verified.
\end{proof}

Here, we have established Proposition \ref{prop:eq_exists} for coincident and progressive peak pricing. We now complete the proof of Theorems \ref{thm:peaks_baseline} and \ref{thm:pp_deterministic} by establishing the final claim:

\begin{claim}
    If $\Delta \leq 0$, then $\mathcal{P} \left( \mathcal{G}^{\rm CP} \right) = \mathcal{P} \left( \mathcal{G}^{\rm PP} \right) = \sum_{t = 1}^T \frac{b^t}{T} + \sum_{i = 1}^N \frac{r_i}{T}.$
\end{claim}

\begin{proof}
To establish (b), observe that taking the limit as $\varepsilon \to 0^+$ with $\bar{a}$ above yields
\begin{equation*}
    \mathcal{P} \left( \mathcal{G}^{\rm CP} \right) \geq b^{t_{\max}} -\frac{\Delta}{T} = \frac{\sum_{t = 1}^T b^t + \sum_{i = 1}^N r_i}{T}.
\end{equation*}
Following similar reasoning as in the previous case, it is readily verified that for any sufficiently small $\varepsilon > 0$, any action profile $\breve{a}$ with consumption
\begin{equation*}
    \frac{\sum_{t = 1}^T b^t + \sum_{i = 1}^N r_i}{T} + \eta
\end{equation*}
in some period $\breve{t}$, where $\eta$ is defined in \eqref{eq:eta_upper_bound}, cannot be an $\varepsilon$-Nash equilibrium; taking the limit as $\varepsilon \to 0^+$ yields that $\mathcal{P} \left( \mathcal{G}^{\rm CP} \right)$ is also upper bounded by $(\sum_{t = 1}^T b^t + \sum_{i = 1}^N r_i)/T$.
\end{proof}

\subsection{Proof of Proposition \ref{prop:progressive_peak_pricing}}\label{sec:proof_my_stochastic}

\subsubsection{Anytime peak pricing (${\rm AP}$)} Under anytime peak pricing, the Players are not affected by the presence of baseline loads on the grid, since $J_i^{\rm AP}$ is not a function of $b$. Hence, it is still optimal for Player $i$ to allocate their consumption evenly over all $T$ periods. Substituting $\bar{a}_i^t = \frac{r_i}{T}$ into \eqref{eq:worst_case_peak_definition} yields
\begin{equation}\label{eq:ca_worst_peak}
    \mathcal{P} \left( \mathcal{G}^{\rm AP} \right) = \bar{b} + \frac{1}{T} \sum_{i = 1}^N r_i.
\end{equation}

\subsubsection{Coincident peak pricing (${\rm CP}$)} Under coincident peak pricing, Player $i$ seeks to solve
\begin{gather*}
    \min_{a_i \in \mathcal{A}_{i}} \, \mathbb{E} \left[ \sum_{t = 1}^T p^t a_i^t + I(\hat{t} = t) \cdot p^{\rm D} a_i^t \right]
    \intertext{which is equivalent to minimizing}
    \sum_{t = 1}^T \mathbb{P}[\hat{t} = t] \cdot a_i^t = \sum_{t = 1}^T \pi^t a_i^t,
\end{gather*}
where
$$\pi^t = \mathbb{P}\left[b^t = \max_{s \in [T]} b^s \right]$$
by \eqref{eq:action_doesnt_matter}. Note that by the continuity of $B$, a tie occurs with probability $0$, so we can neglect the choice of tie-breaking rule. This is a linear program in $a_i$, so an optimal solution for Player $i$ is simply
\begin{equation*}
    \bar{a}_i^t = \begin{cases}
        r_i & t = \min (\argmin_t \pi^t) \\ 
        0 & \text{otherwise.}
    \end{cases}
\end{equation*}
Summing over all Players and taking the maximum over $b \in \text{supp}(B)$ yields
\begin{equation}\label{eq:tx_worst_peak}
    \mathcal{P} \left( \mathcal{G}^{\rm CP} \right) = \bar{b} + \sum_{i = 1}^N r_i.
\end{equation}

\subsubsection{Progressive peak pricing} Under progressive peak pricing, it is readily verified that Player $i$'s optimization problem can be written as
\begin{equation*}
    \min_{a_i \in \mathcal{A}_i} \; \sum_{t = 1}^T \pi^t \left( a_i^t \right)^\delta.
\end{equation*}
Note that if $\pi^{t'} = 0$ for any $t' \in [T]$, then a trivially optimal solution is $\bar{a}_i^{t'} = r_i$ for all $i$, in which case $\mathcal{P}\left( \mathcal{G}^{\rm PP} \right) = \mathcal{P}\left( \mathcal{G}^{\rm CP} \right)$. Otherwise, suppose $\pi^t > 0$ for all $t$. Using $\lambda \in \mathbb{R}_{\geq 0}^T$ to denote the Lagrange multipliers for the nonnegativity constraints and $\nu$ for the electricity consumption constraint, the Lagrangian is given by
\begin{align*}
    L(a_i, \lambda, \nu) &= \nu r_i + \sum_{t = 1}^T \pi^t \left(  a_i^t \right)^\delta - (\lambda^t + \nu) a_i^t .
\end{align*}
Since this problem is strictly convex, we can use the Karush-Kuhn-Tucker (KKT) conditions to identify the unique minimizer. First, stationarity yields
\begin{gather*}
    \frac{\partial L}{\partial \bar{a}_i^t} = \delta \pi^t \left( \bar{a}_i^t \right)^{\delta - 1} - \bar{\lambda}^t - \bar{\nu} = 0 \\
    \implies \bar{a}_i^t = \left( \frac{\bar{\nu} + \bar{\lambda}^t}{ \delta \pi^t} \right)^{\frac{1}{\delta - 1}} . \nonumber
\end{gather*}
By complementary slackness ($\bar{a}_i^t \bar{\lambda}^t = 0$), we have that $\bar{\lambda}^t = 0$ for all $t$. Thus, summing over all $t$ yields
\begin{equation*}
    \sum_{t = 1}^T  \left( \frac{\bar{\nu}}{ \delta \pi^t} \right)^{\frac{1}{\delta - 1}} = r_i \implies \bar{\nu} = \left( \frac{r_i}{\sum_{t = 1}^T (\delta \pi^t)^{-\frac{1}{\delta - 1}}} \right)^{\delta - 1}.
\end{equation*}
Summing over $i$, taking the maximum over $t$, and taking the maximum over $b \in \text{supp}(B)$ yields
\begin{equation}\label{eq:my_worst_peak}
    \mathcal{P} \left( \mathcal{G}^{\rm PP} \right) = \bar{b} + \left( \frac{ \left(\delta \min_t \pi^t \right)^{-\frac{1}{\delta - 1}}}{\sum_{t = 1}^T (\delta \pi^t)^{-\frac{1}{\delta - 1}}} \right) \sum_{i = 1}^N r_i.
\end{equation}
The quantity multiplying $\sum_{i = 1}^N r_i$ is lower and upper bounded by $\frac{1}{T}$ and $1$, respectively, so comparing \eqref{eq:ca_worst_peak}, \eqref{eq:tx_worst_peak}, and \eqref{eq:my_worst_peak} yields \eqref{eq:pp_random_baseline_relation}.

\end{document}